\documentclass[english,preprint,showpacs,preprintnumbers,amsmath,amssymb,floatfix]{revtex4}
\usepackage{graphicx}

\usepackage{bm}
\usepackage{dcolumn}

\begin{document}

\title{Spreading Dynamics of Polymer Nanodroplets in Cylindrical Geometries}

\author{David R. Heine, Gary S. Grest, and Edmund B. Webb III}

\affiliation{Sandia National Laboratories, Albuquerque, New Mexico 87185}

\date{\today}

\begin{abstract}
The spreading of one- and two-component polymer nanodroplets is studied
using molecular dynamics simulation in a cylindrical geometry. The
droplets consist of polymer chains of length 10, 40, and 100 monomers
per chain described by the bead-spring model spreading on a flat surface
with a surface-coupled Langevin thermostat. Each droplet contains
$\sim350\,000$ monomers. The dynamics of the individual components
of each droplet are analyzed and compared to the dynamics of single
component droplets for the spreading rates of the precursor foot and
bulk droplet, the time evolution of the contact angle, and the velocity
distribution inside the droplet. We derive spreading models for the
cylindrical geometry analogous to the kinetic and hydrodynamic models
previously developed for the spherical geometry and show that hydrodynamic
behavior is observed at earlier times for the cylindrical geometry.
The contact radius is predicted to scale like $r(t)\sim t^{1/5}$
from the kinetic model and $r(t)\sim t^{1/7}$ for the hydrodynamic
model in the cylindrical geometry.
\end{abstract}

\maketitle

\section{Introduction}

Practical applications of the spreading of a liquid on a solid are
prevalent in the lubrication, coatings, and printing industries, to
name a few. Knowledge of the rates of spreading and equilibrium configurations
of these systems are crucial for improving their performance. Extensive
experimental, theoretical, and computational work has been undertaken
to better understand the interaction between a liquid and solid in
contact. Most frequently, the liquids studied are oligomers or polymers
in order to remove the influence of evaporation and condensation on
the droplet spreading dynamics.

The total energy dissipation in a spreading droplet can be represented
as a sum of three different components; one due to the hydrodynamic
flow in the bulk of the droplet, one due to the viscous dissipation
in the precursor foot, and one due to the adsorption and desorption
of molecules to the solid surface in the vicinity of the contact line
\cite{G:RMP:85}. Experimental measurements \cite{RCO:Lan:99,RCV:Lan:00}
of microscopic droplets compare well with the hydrodynamic model of
droplet spreading \cite{HS:JCI:71,T:JPD:79,V:FD:76,C:JFM:86}, indicating
that hydrodynamic energy dissipation is an important feature of droplet
spreading. To date, however, simulations \cite{YKB:PRL:91,YKB:PRA:92,NAK:PRL:92,COK:PRL:95,OCK:PRE:96,RBC:Lan:99,BCC:CSA:99,RBC:JPS:99,HGW:PRE:03}
of spherical droplets have been unable to approach the droplet size
and time duration required for hydrodynamic flow to be relevant. 

Although simulations of spreading droplets typically consider a three-dimensional
spreading hemisphere, there are computational advantages for considering
a two-dimensional hemicylinder \cite{RMS:JCP:99,F:PF:03}. The symmetry
along the cylinder axis allows periodic boundary conditions to be
applied in one direction, and a larger droplet radius $r$ can be
simulated with fewer atoms since the droplet volume scales as $r^{2}$
in the cylindrical geometry instead of $r^{3}$ for the spherical
geometry. With larger droplet sizes, this enables us to simulate more
viscous systems by including polymer chains of length $N=100$ for
the first time in droplet spreading simulations. 

It has been claimed that hydrodynamic dissipation is dominant for
small contact angles and non-hydrodynamic dissipation is dominant
for relatively large contact angles \cite{BG:ACI:92}. This is reinforced
by the fact that for spherical droplets, spreading models have a kinetic
dissipation term that is linear in the instantaneous contact radius
while the hydrodynamic dissipation term has a logarithmic dependence
on the instantaneous contact radius \cite{B:Wet:93,RCO:Lan:99,RCV:Lan:00}.
We show here that for a cylindrical geometry, the hydrodynamic dissipation
term is linearly dependent on the contact radius, which suggests that
hydrodynamic flow could contribute to the dissipation at earlier times
for a cylindrical geometry than for the spherical geometry. Our simulations
show that this is indeed the case. We also show that the $r(t)\sim t^{1/10}$
scaling of Tanner's spreading law and the $r(t)\sim t^{1/7}$ prediction
of molecular-kinetic theory for spherical droplets become $r(t)\sim t^{1/7}$
and $r(t)\sim t^{1/5}$, respectively, in the cylindrical geometry.

Even though many of the liquids used in surface wetting applications
are mixtures or suspensions, most of the research has focused on single
component liquids. Some experimental \cite{RAR:MAC:92,SKE:SCI:92,FB:EL:97,BFB:IJE:00}
and theoretical \cite{PIR:JSP:03,YC:JCP:03} work has been done on
binary droplets focusing mainly on the equilibrium behavior. Simulations
of binary droplets containing from $4\,000$ \cite{NA:EL:94,NA:PRE:94}
to $25\,000$ \cite{VRC:Lan:00} monomers have been performed, but
larger system sizes are needed to adequately model the spreading dynamics.

In this paper, we present MD simulations of coarse-grained models
of one- and two-component polymer droplets for chain length $N=10$,
40, and 100. These chain lengths are chosen since they have a very
low vapor pressure and the droplet spreading is not influenced by
vaporization and condensation. We analyze the dynamics of the components
of each droplet and compare the spreading rates of two-component droplets
to their single-component analogues. We derive the equations for the
rate of change of contact angle and radius for the cylindrical geometry
based on kinetic \cite{GLE:TRP:41,B:CAT:68,BH:JCI:69} and hydrodynamic
models \cite{C:JFM:86,SB:JAP:94,RCO:Lan:99} and we use these models
to extract physical parameters for each system.

The paper is organized as follows. Section \ref{sec:Simulation-Details}
describes the details of the molecular dynamics simulations and the
application of the Langevin thermostat to the monomers near the substrate.
It also describes the methods used to analyze the simulation results.
Section \ref{sec:Cylindrical-geometry-droplet} presents the droplet
spreading models for the cylindrical geometry. Section \ref{sec:Results}
compares the spreading behavior of monodispersed droplets in the spherical
and cylindrical geometry, droplets of different chain lengths, and
binary mixtures. The velocity distributions of both homogeneous and
binary droplets are analyzed in Section \ref{sec:Velocity-distribution}
and conclusions are presented in Section \ref{sec:Conclusions}.

\section{Simulation Details\label{sec:Simulation-Details}}

\subsection{Potentials and Thermostat}

Molecular dynamics (MD) simulations are performed using a coarse-grained
model for the polymer chains in which the polymer is represented by
spherical beads of mass $m$ attached by springs. We use a truncated
Lennard-Jones (LJ) potential to describe the interaction between the
monomers. The LJ potential is given by

\begin{equation}
U_{LJ}(R)=\left\{ \begin{array}{rl}
4\varepsilon\left[\left(\frac{\sigma}{r}\right)^{12}-\left(\frac{\sigma}{r}\right)^{6}\right] & r\leq r_{c}\\
0 & r>r_{c}\end{array}\right.\label{eq:ljcut}\end{equation}
 where $\varepsilon$ and $\sigma$ are the LJ units of energy and
length and the cutoff is set to $r_{c}=2.5\:\sigma$. The monomer-monomer
interaction $\varepsilon$ is used as the reference and all monomers
have the same diameter $\sigma$. Although in this paper for the binary
mixtures we vary only the chain length, in future work we will vary
the relative interaction strength. For bonded monomers, we apply an
additional potential where each bond is described by the finite extensible
nonlinear elastic (FENE) potential \cite{KG:JCP:90},\begin{equation}
U_{FENE}(r)=\left\{ \begin{array}{rl}
\frac{-k}{2}R_{0}^{2}\ln\left[1-\left(\frac{r}{R_{0}}\right)^{2}\right] & r\leq R_{0}\\
\infty & r>R_{0}\end{array}\right.,\label{eq:fene}\end{equation}
 with $k=30\:\varepsilon$ and $R_{0}=1.5\:\sigma$.

Droplets composed of polymer chains of length N=10, 40, 100, or 10/40
and 10/100 mixtures of equal mole fraction of monomers are generated
by equilibrating a melt of the polymer containing $500\,000$ monomers
at $P\simeq0$ between two parallel plates in the $z$ direction with
periodic boundary conditions in the other two directions. The distance
between the plates $L_{z}\simeq90\,\sigma$. For the cylindrical geometry,
the width of the simulation cell in the $y$ direction is chosen to
be wide enough such that there are no interactions between a chain
and its periodic image. The larger the width, the better are the statistical
averaging of contact angle and radius measurements of the droplets.
We found that both $L_{y}=40$ and $60\:\sigma$ give results with
reasonable uncertainty in these measurements. For the spherical droplet,
the dimensions $L_{x}=L_{y}$. The cylindrical droplets were constructed
by removing all chains with centers outside of a hemicylinder of radius
$R_{0}=80\:\sigma$ in the xz plane and $L_{y}=40\,\sigma$, which
resulted in droplets containing $\sim350\,000$ monomers. For $R_{0}=50\:\sigma$
and $L_{y}=60\,\sigma$, the droplets contained $\sim200\,000$ monomers.
Hemispherical droplets were constructed in a similar manner, with
initial radii $\sim48\:\sigma$, resulting in a droplet also of $\sim200\,000$
monomers. The droplets were then placed above a substrate which initially
has an interaction strength chosen so that the droplet equilibrates
with a contact angle near $90^{o}$. This is necessary since the method
of construction of the drop leaves some segments extending into the
vapor phase. These dangling chain segments quickly coalesce with the
droplet after a short equilibration run. Hemispheres and hemicylinders
were chosen over spheres and cylinders to avoid the substantial simulation
time required for the isotropic droplet to transition to a cap geometry
\cite{HGW:PRE:03}. All of the droplets studied here are large enough
to avoid the equilibrium contact angle dependence on droplet size
observed for smaller system sizes \cite{HGW:PRE:03}.

The substrate is modeled as a flat surface since it was found previously
\cite{HGW:PRE:03} that with the proper choice of thermostat, the
simulations using a flat surface exhibit the same behavior as a realistic
atomic substrate. Since simulating a realistic substrate requires
several times the total number of atoms in the simulation, using the
flat surface greatly improves the computational efficiency. The interactions
between the surface and the monomers in the droplet at a distance
$z$ from the surface are modeled using an integrated LJ potential,

\begin{equation}
U_{LJ}^{wall}(z)=\left\{ \begin{array}{rl}
\frac{2\pi\varepsilon_{w}}{3}\left[\frac{2}{15}\left(\frac{\sigma}{z}\right)^{9}-\left(\frac{\sigma}{z}\right)^{3}\right] & z\leq z_{c}\\
0 & z>z_{c}\end{array}\right.\label{eq:ljwall}\end{equation}
 with $z_{c}=2.2\sigma.$

We apply the Langevin thermostat to provide a realistic representation
of the transfer of energy in the droplet. The Langevin thermostat
simulates a heat bath by adding Gaussian white noise and friction
terms to the equation of motion,

\begin{equation}
m_{i}\mathbf{\ddot{r}}_{i}=-\Delta U_{i}-m_{i}\gamma_{L}\dot{\mathbf{r}_{i}}+\mathbf{W}_{i}(t),\label{eq:lang}\end{equation}
 where $m_{i}$ is the mass of monomer $i$, $\gamma_{L}$ is the
friction parameter for the Langevin thermostat, $-\Delta U_{i}$ is
the force acting on monomer $i$ due to the potentials defined above,
and $\mathbf{W}_{i}(t)$ is a Gaussian white noise term such that

\begin{equation}
\left\langle \mathbf{W}_{i}(t)\cdot\mathbf{W}_{j}(t')\right\rangle =6k_{B}Tm_{i}\gamma_{L}\delta_{ij}\delta\left(t-t'\right).\label{eq:whitenoise}\end{equation}
Coupling all of the monomers to the Langevin thermostat would have
the unphysical effect of screening the hydrodynamic interactions in
the droplet and not damping the monomers near the surface stronger
than those in the bulk. To overcome this, we use a Langevin coupling
term with a damping rate that decreases exponentially away from the
substrate \cite{BP:PRE:01}. We choose the form

\begin{equation}
\gamma_{L}(z)=\gamma_{L}^{s}\exp\left(\sigma-z\right)\label{eq:expdamp}\end{equation}
 where $\gamma_{L}^{s}$ is the surface Langevin coupling and $z$
is the distance from the substrate. We generally use values of $\gamma_{L}^{s}=10.0\,\tau^{-1}$
and $3.0\,\tau^{-1}$ for $\varepsilon_{w}=2.0\,\varepsilon$ and
$3.0\,\varepsilon$, respectively, based on earlier work \cite{HGW:PRE:03}
matching the diffusion constant of the precursor foot for flat and
atomistic substrates. The larger $\gamma_{L}^{s}$ corresponds to
an atomistic substrate with larger corrugation and hence larger dissipation
and slower diffusion near the substrate.

The equations of motion are integrated using a velocity-Verlet algorithm.
We use a time step of $\Delta t=0.01\;\tau$ where $\tau=\sigma\left(\frac{m}{\varepsilon}\right)^{1/2}$.
The simulations are performed at a temperature $T=\varepsilon/k_{B}$
using the \textsc{lammps} code \cite{P:JCP:95}. Most of the simulations
were run on $64$ to $100$ processors of Sandia's ICC Intel Xeon
cluster. One million steps for a wetting drop of $350\,000$ monomers
takes 24 to 86 hours on 64 processors depending on the radius of the
droplet.

\subsection{Analysis details}

For all of the simulations presented here, we extract the instantaneous
contact radius $r(t)$ and contact angle $\theta(t)$ every $400\,\tau$.
The contact radius is calculated by defining a one-dimensional radial
distribution function, $g(r)=\rho(r)/\rho$, based on every monomer
within $1.5\:\sigma$ of the surface. The local density at a distance
$r$ from the center of mass of the droplet is \begin{equation}
\rho(r)=\frac{N(r)}{2L\Delta r}\label{eq:rdf}\end{equation}
 where $N(r)$ is the number of monomers at a distance between $r$
and $r+\Delta r$ from the center of mass, $L$ is the width of the
simulation cell in the periodic direction, and $\rho$ is the integral
of $\rho(r)$ over the entire surface. The contact radius is defined
as the distance $r$ at which $g(r)=0.98$. This approach provides
a robust measure of the radius at any point during the spreading simulation.
The same calculation is used to obtain the droplet radius for ten
slices of the droplet at incremental heights every $1.5\:\sigma$
from the surface. A line is fit to the resulting points and the instantaneous
contact angle is determined from the slope of the line. For simulations
that exhibit a precursor foot, the monomers within $4.5\:\sigma$
of the surface are ignored in the contact angle calculation.

\begin{figure}

\includegraphics[%
  clip,
  width=8.5cm]{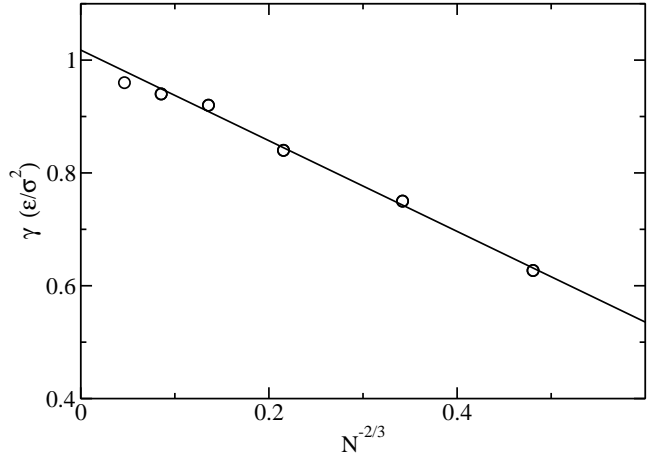}
\caption{\label{cap:surftens} Molecular weight dependence of the liquid/vapor
surface tension for $T=\varepsilon/k_{B}$. The solid line is a fit
to the experimentally observed $N^{-2/3}$ dependence.}

\end{figure}

Fitting the spreading models to the contact angle and contact radius
data requires knowledge of both the surface tension and viscosity.
Both involve separate simulations of bulk polymers. The surface tension
is obtained by simulating the polymer melt in a slab geometry so that
there are two surfaces perpendicular to the z
 direction. For the $N\geq40$ system, the melt contains $200\,000$
monomers and each surface has a cross-sectional area of $4900\,\sigma^{2}$.
For shorter chain lengths, the melts contain $100\,000$ monomers
and each surface has a cross-sectional area of $2500\,\sigma^{2}$.
After the system equilibrates, the surface tension is calculated from
the parallel and perpendicular components of the pressure tensor via
\cite{NBB:JCP:88}

\begin{equation}
\gamma=\frac{1}{2}\int_{0}^{L_{z}}\left[p_{\perp}\left(z\right)-p_{\parallel}\left(z\right)\right]dz.\label{eq:surftens}\end{equation}
The driving force for the spontaneous spreading of a droplet is the
difference in surface tension at each interface. Since the liquid/vapor
surface tension $\gamma$ is dependent on the chain length, the spreading
rate is as well. Figure \ref{cap:surftens} shows a plot of $\gamma$
for droplets of several chain lengths obtained from MD simulation.
The data fits the experimentally \cite{LG:JCIS:75} observed molecular
weight dependence $\gamma\sim N^{-2/3}$ very well and provides a
means to both extrapolate values of $\gamma$ for large molecular
weights and estimate the change in spreading rate for different chain
lengths.

\begin{figure}

\includegraphics[%
  clip,
  width=8.5cm]{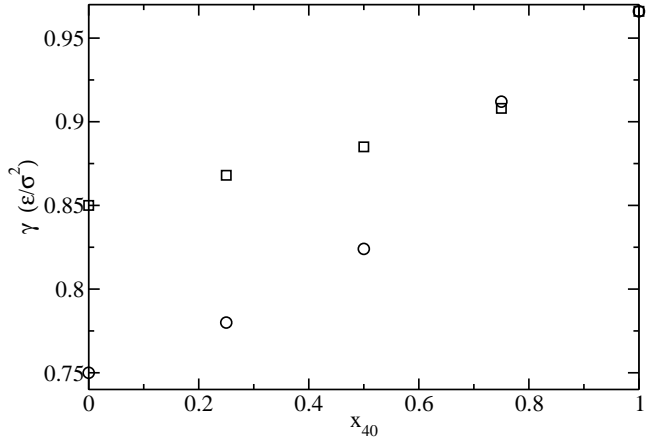}
\caption{\label{cap:surftensblend} Surface tension of binary blends of N=40
polymers with N=5 polymers ($\circ$) and N=10 polymers ($\square$)
as a function of the bulk mole fraction $x_{40}$ of monomers on N=40
chains.}

\end{figure}

The surface tensions of the binary droplets are also obtained from
mixtures of the two components. To determine the composition dependence
of the surface tension, we equilibrate blends of $N=40$ with $N=5$
and with $N=10$ at three blend compositions as shown in Fig. \ref{cap:surftensblend}
and $N=100$ with $N=10$ at a composition $x_{100}=0.5$. This allows
us to compare the cases where there is a large (5/40 system) or a
moderate difference (10/40 system) in the surface tension of the pure
components. For the binary systems, the mixtures contain $200\,000$
monomers and each surface has a cross-sectional area of $4900\,\sigma^{2}$.
Note that the surface tension shown in Fig. \ref{cap:surftensblend}
is not a simple mean field (i.e. linear) function of the monomer fraction
as the fully equilibrated surface composition consists of almost fully
shorter chains. 

The viscosity is calculated from the equilibrium fluctuations of the
off-diagonal components of the stress tensor \cite{AT:CSL:87} obtained
from polymer melt simulations at $T=\varepsilon/k_{B}$ with the bulk
pressure $P\simeq0$ without tail corrections. We do not include the
tail corrections to the pressure in order to match the system of the
spreading droplet. These simulations are run up to $84,000\:\tau$.
The autocorrelation function of each off-diagonal component of the
stress tensor is calculated using the Numerical Recipes routine \textsc{correl}
\cite{PTV:NR:92}. The autocorrelation functions are averaged to improve
statistical uncertainty. From this, the viscosity can be calculated
using \cite{AT:CSL:87}\begin{equation}
\eta=\frac{V}{k_{B}T}\int_{0}^{\infty}dt\left\langle \sigma_{\alpha\beta}(t)\sigma_{\alpha\beta}(0)\right\rangle \label{eq:visc}\end{equation}
 where $V$ is the system volume and $\sigma_{\alpha\beta}(t)$ is
the $\alpha\beta$ component of the stress tensor at time $t$. The
results for $\gamma$ and $\eta$ are also summarized in Table \ref{cap:paramtable}. 

\begin{table}

\caption{\label{cap:paramtable} Bulk properties of bead-spring chains obtained
from MD simulation and model fit parameters for $T=\varepsilon/k_{B}$,
$P\simeq0.$}

\begin{tabular}{|c|c|c|c|c|c|c|c|c|c|c|c|c|}
\hline 
&
&
&
&
&
&
Kinetic&
Hydro&
\multicolumn{2}{c|}{Combined}&
&
&
\tabularnewline
N&
$\varepsilon_{w}\,\left(\varepsilon\right)$&
$\gamma_{L}^{s}\,\left(\tau^{-1}\right)$&
$\rho\,\left(\sigma^{-3}\right)$&
 $\gamma\,\left(\varepsilon/\sigma^{2}\right)$&
 $\eta\,\left(m/\tau\sigma\right)$&
$\zeta_{0}\:\left(\frac{m}{\tau\sigma}\right)$&
$a\,\left(\sigma\right)$&
$\zeta_{0}\:\left(\frac{m}{\tau\sigma}\right)$&
$a\,\left(\sigma\right)$&
$\chi_{kin}^{2}$&
$\chi_{hydro}^{2}$&
$\chi_{comb}^{2}$\tabularnewline
\hline
10&
2.0&
10.0&
0.869&
0.84$\pm$0.02&
11.1$\pm$0.4&
72.2&
30.3&
124&
190&
.0014&
.0049&
.0012\tabularnewline
\hline 
10&
3.0&
3.0&
0.869&
0.84$\pm$0.02&
11.1$\pm$0.4&
37.4&
60.4&
63.2&
147&
.0029&
.0045&
.0012\tabularnewline
\hline 
40&
2.0&
10.0&
0.886&
0.94$\pm$0.02&
41.7$\pm$1.4&
141&
58.2&
339&
181&
.0008&
.0066&
.0011\tabularnewline
\hline 
40&
3.0&
3.0&
0.886&
0.94$\pm$0.02&
41.7$\pm$1.4&
59.9&
73.6&
160&
140&
.0037&
.011&
.0018\tabularnewline
\hline 
100&
2.0&
10.0&
0.892&
0.96$\pm$0.02&
132$\pm$2&
180&
41.2&
155&
51.8&
.0015&
.0009&
.0009\tabularnewline
\hline 
100&
3.0&
3.0&
0.892&
0.96$\pm$0.02&
132$\pm$2&
82.4&
70.7&
417&
124&
.0057&
.013&
.0019\tabularnewline
\hline 
100&
2.0&
3.0&
0.892&
0.96$\pm$0.02&
132$\pm$2&
105&
65.0&
678&
155&
.0012&
.016&
.0022\tabularnewline
\hline 
100&
3.0&
10.0&
0.892&
0.96$\pm$0.02&
132$\pm$2&
167&
61.2&
126&
77.3&
.0057&
.0005&
.0004\tabularnewline
\hline
\end{tabular}
\end{table}

The viscosity of each blend is obtained in the same manner as the
pure components. The surface tension and viscosity for each blend
is given in Table \ref{cap:binarytable}. The surface tensions of
the mixtures are closer to that of the shorter chains since they dominate
the liquid/vapor interface in the equilibrated system. However, the
viscosity of the mixture is more strongly influenced by the longer
chains.

\begin{table}

\caption{\label{cap:binarytable} Bulk properties and model fit parameters
for binary droplets.}

\begin{tabular}{|c|c|c|c|c|c|c|c|c|c|c|c|c|}
\hline 
&
&
&
&
&
&
Kinetic&
Hydro&
\multicolumn{2}{c|}{Combined}&
&
&
\tabularnewline
N&
$\varepsilon_{w}\,\left(\varepsilon\right)$&
$\gamma_{L}^{s}\,\left(\tau^{-1}\right)$&
$\rho\,\left(\sigma^{-3}\right)$&
$\gamma$ ($\varepsilon/\sigma^{2})$&
$\eta$ (m/$\tau\sigma$)&
$\zeta_{0}\:\left(\frac{m}{\tau\sigma}\right)$&
a$\left(\sigma\right)$&
$\zeta_{0}\:\left(\frac{m}{\tau\sigma}\right)$&
a$\left(\sigma\right)$&
$\chi_{kin}^{2}$&
$\chi_{hydro}^{2}$&
$\chi_{comb}^{2}$\tabularnewline
\hline
10/40&
2.0&
10.0&
0.8800&
0.885$\pm$0.02&
34.8$\pm$1.4&
109&
58.9&
259&
178&
.0009&
.0059&
.0006\tabularnewline
\hline 
10/40&
3.0&
3.0&
0.8800&
0.885$\pm$0.02&
34.8$\pm$1.4&
45.5&
73.9&
168&
160&
.0027&
.015&
.0024\tabularnewline
\hline 
10/100&
3.0&
3.0&
0.8830&
0.90$\pm$0.02&
67.2$\pm$2.4&
52.8&
75.7&
158&
127&
.0061&
.015&
.0023\tabularnewline
\hline
\end{tabular}
\end{table}

\section{Cylindrical geometry droplet spreading models\label{sec:Cylindrical-geometry-droplet}}

\begin{figure}

\includegraphics[%
  clip,
  width=8.5cm]{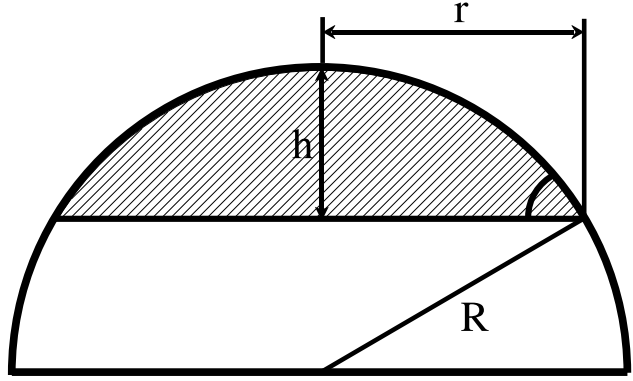}
\caption{\label{cap:capdiag} Diagram of the cylindrical cap.}

\end{figure}

The droplet is modeled as a cylindrical cap as shown by the hatched
region in Figure \ref{cap:capdiag}. Here, the droplet volume is defined
as the cap of height $h$ and width $2r$ of the cylinder with radius
$R$ and length $L$. The cap height can be expressed in terms of
the contact angle, $\theta$, as

\begin{equation}
h=r\frac{1-\cos\left(\theta\right)}{\sin\left(\theta\right)}.\label{eq:height}\end{equation}
Using the definition for the area of a circular segment, $A=\frac{1}{2}R^{2}\left(2\theta-\sin2\theta\right)$,
the radius of the cap can be written in terms of $\theta$ and differentiated
to give

\begin{equation}
\frac{dr}{dt}=\left(\frac{A}{\theta-\sin\theta\cos\theta}\right)^{1/2}\left(\cos\theta-\frac{\sin^{3}\theta}{\theta-\sin\theta\cos\theta}\right)\frac{d\theta}{dt}.\label{eq:drdt}\end{equation}

The free energy is determined by integrating the surface tensions
of the liquid/vapor, solid/vapor and solid/liquid interfaces over
the areas of each interface. For the cylindrical cap geometry shown
in Fig. \ref{cap:capdiag}, this is given by

\begin{equation}
F\left\{ r(t)\right\} =2r(t)L\left(\gamma_{SL}-\gamma_{SV}\right)+2\gamma L\int_{0}^{r(t)}dx\left[1+\left(\frac{\partial h'(x,t)}{\partial x}\right)^{2}\right]^{1/2}\label{eq:F}\end{equation}
where $\gamma$, $\gamma_{SV}$, and $\gamma_{SL}$ are the liquid/vapor,
solid/vapor, and solid/liquid surface tensions, respectively. The
height of the cylindrical cap in terms of the cap dimensions is given
by

\begin{equation}
h'(x,t)=\frac{r(t)}{\sin\theta}\left[1-\left(\frac{x^{2}\sin^{2}\theta}{r(t)^{2}}\right)^{-1/2}-\cos\theta\right].\label{eq:h}\end{equation}
Combining Eqs. \ref{eq:F} and \ref{eq:h} and differentiating gives

\begin{equation}
\frac{\partial F\left\{ r(t)\right\} }{\partial r(t)}=2L\gamma\left(\frac{\theta}{\sin\theta}-\frac{\theta_{0}}{\sin\theta_{0}}\right).\label{eq:F2}\end{equation}

Using the standard mechanical description of dissipative system dynamics,
the dissipation function can be represented as 

\begin{equation}
\frac{\partial T\left\{ r(t);\dot{r}(t)\right\} }{\partial\dot{r}(t)}=\frac{\partial F\left\{ r(t)\right\} }{\partial r(t)}\label{eq:dTdr}\end{equation}
where $T$ is the dissipation function \cite{G:RMP:85,G:LI:90}, which
we consider to be composed of a kinetic component $T\dot{\sum}_{l}$
and a hydrodynamic component $T\dot{\sum}_{w}$. The kinetic dissipation
term, due to molecular adsorption near the contact line, follows the
kinetic model introduced by Eyring and coworkers \cite{GLE:TRP:41}
and applied to spreading of a spherical droplet by Blake and Haynes
\cite{B:CAT:68,BH:JCI:69}. In the kinetic model, the liquid molecules
jump between surface sites separated by a distance $\lambda$ with
a frequency $K$. For the spreading cylinder, the velocity of the
contact line to first order is obtained from Eq. \ref{eq:F2} as

\begin{equation}
\dot{r}(t)=\frac{2\gamma}{\zeta_{0}}\left(\frac{\theta}{\sin\theta}-\frac{\theta_{0}}{\sin\theta_{0}}\right)\label{eq:Blakevel}\end{equation}
where the friction coefficient $\zeta_{0}\equiv\frac{\Delta nk_{B}T}{K\lambda}$.
Here, $\Delta n$ is the density of sites on the solid surface. Combining
Eqs. \ref{eq:F2}-\ref{eq:Blakevel} we find that the dissipation
term due to the surface kinetics is 

\begin{equation}
T\dot{\sum}_{l}=\zeta_{0}\dot{r}(t)^{2}L/2\label{eq:Tl}\end{equation}

\begin{figure}

\includegraphics[%
  clip,
  width=8.5cm]{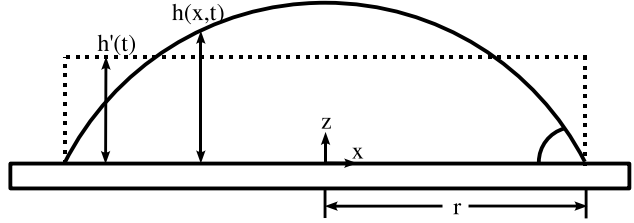}
\caption{\label{cap:hydrodiag} Rectangular representation of the cylindrical
cap.}

\end{figure}

The hydrodynamic dissipation term for the spreading droplet is obtained
by solving the equations of motion and continuity. For the spherical
droplet, Seaver and Berg \cite{SB:JAP:94} found that approximating
the spherical cap as a cylindrical disk of the same volume gave results
that differed from the rigorous derivation by Cox \cite{C:JFM:86}
only by insignificant numerical factors. We apply the same approximation
here, treating the hydrodynamics of the cylinder as identical to that
of a rectangular box as shown in Fig. \ref{cap:hydrodiag}. Following
de Ruijter \textit{et al}. \cite{RCO:Lan:99}, we set the velocity
of the upper part of the leading edge to the droplet spreading velocity,
$v_{x}\left[x=r(t),z=h'\right]=\dot{r}(t)$. With this boundary condition,
the velocity profile is simply \begin{equation}
v_{x}(x,z)=\frac{z}{h'(x,t)}\dot{r}(t).\label{eq:vx}\end{equation}
 The hydrodynamic dissipation $\Sigma_{w}$ is defined as \begin{equation}
T\sum_{w}=\eta\int_{V}dV\left(\frac{\partial v_{x}}{\partial z}\right)^{2}.\label{eq:Twdef}\end{equation}
 Combining Eqs \ref{eq:vx} and \ref{eq:Twdef} and integrating gives

\begin{equation}
T\sum_{w}=L\eta\dot{r}(t)^{2}\left(r(t)-a\right)/h'\label{eq:Tw}\end{equation}
where the parameter $a$ has the same meaning as in \cite{RCO:Lan:99},
it is a minimum radius cutoff applied to avoid the singularity in
the velocity at the z axis. Note that the logarithmic dependence of
$T\sum_{w}$ on $r(t)$ for the case of a spreading sphere \cite{RCO:Lan:99}
becomes a linear dependence for the case of a spreading cylinder.
Since the rectangular box and the cylindrical cap have the same volume,
we can rewrite Eq. \ref{eq:Tw} in terms of the cylinder dimensions

\begin{equation}
T\sum_{w}=2L\eta\dot{r}(t)^{2}\frac{\sin^{2}\theta\left(r(t)-a\right)}{r(t)\left(\theta-\sin\theta\cos\theta\right)}.\label{eq:Tw2}\end{equation}
For the spherical geometry, the hydrodynamic dissipation term has
been derived previously \cite{RCO:Lan:99},

\begin{equation}
T\sum_{w}=6\pi r(t)\eta\phi\left[\theta(t)\right]\dot{r}(t)^{2}\ln\left[r(t)/a\right]\label{eq:dishydrosph}\end{equation}
where $\phi\left[\theta(t)\right]$ is defined as

\begin{equation}
\phi\left[\theta(t)\right]=\frac{\left[1+\cos\theta(t)\right]\sin\theta(t)}{\left[1-\cos\theta(t)\right]\left[2+\cos\theta(t)\right]}.\label{eq:combphi}\end{equation}

We construct a combined kinetic and hydrodynamic model in a manner
analogous to de Ruijter \textit{et al}. by combining Eqs. \ref{eq:F2}-\ref{eq:Tl}
and \ref{eq:Tw2}:

\begin{equation}
\dot{r}(t)=\frac{\gamma}{\frac{\zeta_{0}}{2}+\frac{2\eta\left(r(t)-a\right)\sin^{2}\theta}{r(t)\left(\theta-\sin\theta\cos\theta\right)}}\left(\frac{\theta}{\sin\theta}-\frac{\theta_{0}}{\sin\theta_{0}}\right)\label{eq:combined}\end{equation}
Rewriting this in terms of the contact angle $\theta$ using Eq. \ref{eq:drdt}
gives

\begin{equation}
\frac{d\theta}{dt}=\left(\frac{\theta-\sin\theta\cos\theta}{A}\right)^{1/2}\left(\cos\theta-\frac{\sin^{3}\theta}{\theta-\sin\theta\cos\theta}\right)^{-1}\frac{\gamma\left(\frac{\theta}{\sin\theta}-\frac{\theta_{0}}{\sin\theta_{0}}\right)}{\frac{\zeta_{0}}{2}+\frac{2\eta\left(r(t)-a\right)\sin^{2}\theta}{r(t)\left(\theta-\sin\theta\cos\theta\right)}}.\label{eq:combined2}\end{equation}
This can be compared to the analogous expressions for a spherical
droplet \cite{RCO:Lan:99,RCV:Lan:00}:

\begin{equation}
\dot{r}(t)=\frac{\gamma\left(\cos\theta_{0}-\cos\theta\right)}{\zeta_{0}+6\eta\phi\left[\theta(t)\right]\ln\left(\frac{r(t)}{a}\right)}\label{eq:combined3}\end{equation}

\begin{equation}
\frac{d\theta}{dt}=-\left(\frac{\pi}{3V}\right)^{1/3}\frac{\left(2-3\cos\theta+\cos^{3}\theta\right)^{4/3}}{\left(1-\cos\theta\right)^{2}}\frac{\gamma\left(\cos\theta_{0}-\cos\theta\right)}{\zeta_{0}+6\eta\phi\left[\theta(t)\right]\ln\left(\frac{r(t)}{a}\right)}.\label{eq:combined4}\end{equation}
The kinetic model is obtained by setting $\eta=0$ in Eqs. \ref{eq:combined}
through \ref{eq:combined4} and the hydrodynamic model is obtained
by setting $\zeta_{0}=0$. For the kinetic model, the asymptotic solutions
of Eqs. \ref{eq:combined} and \ref{eq:combined2} give 

\begin{equation}
\theta(t)\sim\left(2A\right)^{1/5}\left(\frac{5\gamma t}{6\zeta_{0}}\right)^{-2/5},\label{eq:asympkq}\end{equation}

\begin{equation}
r(t)\sim2\left(2A\right)^{2/5}\left(\frac{5\gamma t}{6\zeta_{0}}\right)^{1/5}\label{eq:asympkr}\end{equation}
as compared to $\theta(t)\sim t^{-3/7}$ and $r(t)\sim t^{-1/7}$
for the spherical geometry. Similarly, the asymptotic solutions for
the hydrodynamic model give

\begin{equation}
\theta(t)\sim\left(2A\right)^{1/7}\left(\frac{7\gamma t}{48\eta}\right)^{-2/7},\label{eq:asymphq}\end{equation}

\begin{equation}
r(t)\sim2\left(2A\right)^{3/7}\left(\frac{7\gamma t}{48\eta}\right)^{1/7}\label{eq:asymphr}\end{equation}
as compared to $\theta(t)\sim t^{-3/10}$ and $r(t)\sim t^{-1/10}$
for the spherical geometry.

\section{Results\label{sec:Results}}

\subsection{Comparison to spherical geometry}

\begin{figure}

\includegraphics[%
  clip,
  width=8.5cm]{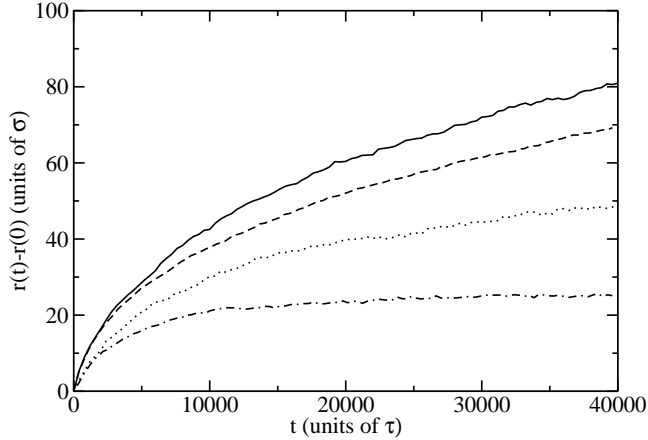}
\caption{\label{cap:sphmodelrad} Radius of the precursor foot for the cylindrical
(solid line) and spherical (dashed line) geometries and of the bulk
droplet for the cylindrical (dotted line) and spherical (dash-dotted
line) geometries. Both cylindrical and spherical droplets consist
of $20\,000$ polymers of length $N=10$. The substrate interaction
strength $\varepsilon_{w}=2.0\,\varepsilon$ is in the fully wetting
regime for $N=10$, $\gamma_{L}^{s}=10.0\,\tau^{-1}$.}

\end{figure}

For wetting droplets, the spreading is characterized by the formation
of a precursor foot of monolayer thickness that advances ahead of
the bulk of the droplet. The bulk region of the droplet follows the
precursor foot at a slower spreading rate. This is demonstrated in
Fig. \ref{cap:sphmodelrad} where the contact radius of the foot and
bulk regions are plotted as a function of time for both the cylindrical
and spherical geometries. These droplets contain $20\,000$ chains
of $N=10$ with a substrate interaction strength $\varepsilon_{w}=2.0\varepsilon$,
which is in the fully wetting regime for $N=10$. For the otherwise
identical systems, the radii of both regions of the cylindrical droplet
increase faster than those for the spherical droplet. This is a consequence
of the droplet spreading in one dimension in the cylindrical geometry
and two dimensions in the spherical geometry. The precursor foot grows
diffusively, $r^{2}(t)\sim t$, in both cases. We return to further
discussion of the time dependence of $r(t)$ in Sec. \ref{sub:Binary-droplets}.

\begin{figure}

\includegraphics[%
  clip,
  width=8.5cm]{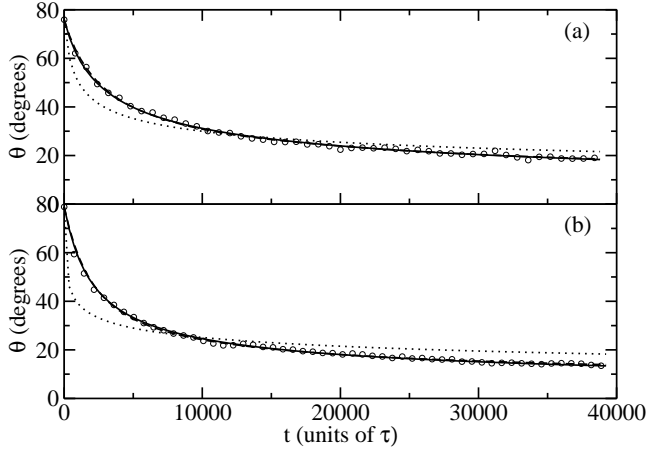}
\caption{\label{cap:2d3dcomp} (a) Fit of the kinetic (solid line), hydrodynamic
(dotted line), and combined (dashed line) models to the contact angle
data for the cylindrical geometry obtained from MD simulation ($\circ$).
(b) Model fits for the equivalent droplet in the spherical geometry.
Both cylindrical and spherical droplets consist of $20\,000$ polymers
of length $N=10$. The substrate interaction strength $\varepsilon_{w}=2.0\,\varepsilon$
is in the fully wetting regime for $N=10$, $\gamma_{L}^{s}=10.0\,\tau^{-1}$.}

\end{figure}

Figure \ref{cap:2d3dcomp} shows the time dependence of the contact
angle for the same system. For the contact angle, the cylindrical
and spherical geometries show a comparable spreading rate. Since the
droplet volume scales as $r^{3}$ in the spherical geometry and $r^{2}$
in the cylindrical geometry, we favor the cylindrical geometry in
order to simulate effectively larger droplets with the same number
of monomers.

Fits to the kinetic, hydrodynamic, and combined models are performed
by taking initial guess values for the independent parameters and
integrating the expression for $d\theta/dt$ defined in Eqs. \ref{eq:combined2}
and \ref{eq:combined4} for the cylindrical and spherical droplets,
respectively. As these data are in the completely wetting regime,
the equilibrium contact angle is fixed at $\theta_{0}=0^{o}$. The
integration uses the fourth-order Runge-Kutta method to generate a
set of data, $\theta_{calc}(t)$. The parameters are varied using
the downhill simplex method \cite{PTV:NR:92} until the difference
between the model and simulation data, $\left|\theta_{calc}(t)-\theta(t)\right|/\theta(t)$,
is minimized. The error reported for each model is calculated as \begin{equation}
\chi^{2}=\frac{1}{{\cal N}}\sum_{i=1}^{{\cal N}}\frac{\left|\theta_{calc}(t)-\theta(t)\right|^{2}}{\theta(t)}\label{eq:error}\end{equation}
 where ${\cal N}$ is the number of data points in each set of data.

For the data shown in Fig. \ref{cap:2d3dcomp}, the hydrodynamic model
provides a more accurate, though still only approximate, fit to the
data in the cylindrical geometry. The hydrodynamic cutoff $a\simeq38.1\,\sigma$
for the spherical geometry and $25.3\,\sigma$ for the cylindrical
geometry, indicating stronger hydrodynamic dissipation in the cylindrical
geometry. For comparison, the friction coefficient obtained from the
kinetic model $\zeta_{0}=56.3\, m/\tau\sigma$ for the spherical geometry
and $56.4\, m/\tau\sigma$ for the cylindrical geometry, are in excellent
agreement.

\subsection{Chain Length Dependence }

\begin{figure}

\includegraphics[%
  clip,
  width=8.5cm]{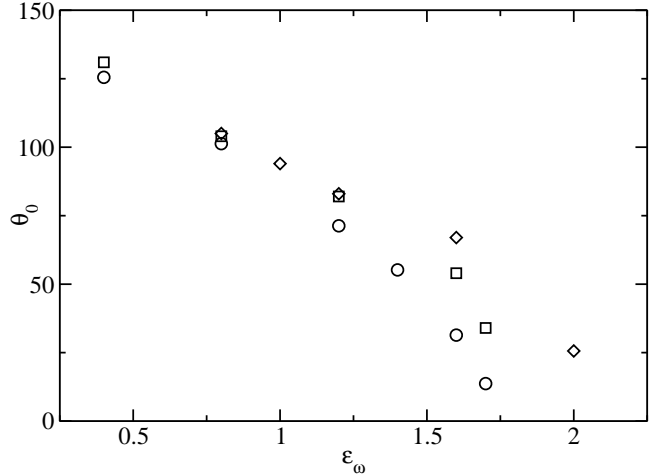}
\caption{\label{cap:eqangle} Equilibrium contact angle as a function of surface
interaction strength for polymer droplets composed of $N=10$ ($\circ$),
$N=40$ ($\square$), and $N=100$ ($\diamond$) monomers per polymer.}

\end{figure}

The equilibrium contact angles for nonwetting droplets are plotted
as a function of the surface interaction strength in Fig. \ref{cap:eqangle}.
The variation of the surface tension with chain length, shown in Fig.
\ref{cap:surftens}, causes a shift in the wetting transition in terms
of the surface interaction strength. The contact angles for the $N=10$
and $N=40$ droplets are taken from earlier work \cite{HGW:PRE:03}
on spherical droplets. The droplets are large enough to eliminate
any equilibrium contact angle dependence on the droplet size \cite{HGW:PRE:03}.
Contact angles for the $N=100$ droplets are from simulations containing
$355\,000$ total monomers. Although the chain length dependence is
weak for small $\varepsilon_{w}$, the wetting transition is shifted
to higher $\varepsilon_{w}$ for larger chain lengths due to the increase
in the liquid vapor surface tension. The transition occurs near $\varepsilon_{w}^{c}\simeq1.75\:\varepsilon$
for $N=10$ droplets and increases to about $\varepsilon_{w}^{c}\simeq2.25\:\varepsilon$
for $N=100$ droplets. Hence, results shown for $\varepsilon_{w}=3.0\:\varepsilon$
are in the completely wetting regime for all chain lengths while those
for $\varepsilon_{w}=2.0\:\varepsilon$ are in the completely wetting
regime only for $N=10$ and $N=40$. From Fig. \ref{cap:eqangle},
the equilibrium contact angle for $N=100$, $\varepsilon_{w}=2.0\:\varepsilon$
is $\theta_{0}\simeq26^{o}$. The equilibrium contact angle extracted
from the kinetic model fit for this droplet is $\theta_{0}\simeq28^{o}$.
For droplets in the completely wetting regime, the models are fit
using an equilibrium contact angle fixed at $\theta_{0}=0^{o}$.

\begin{figure}

\includegraphics[%
  clip,
  width=8.5cm,
  keepaspectratio]{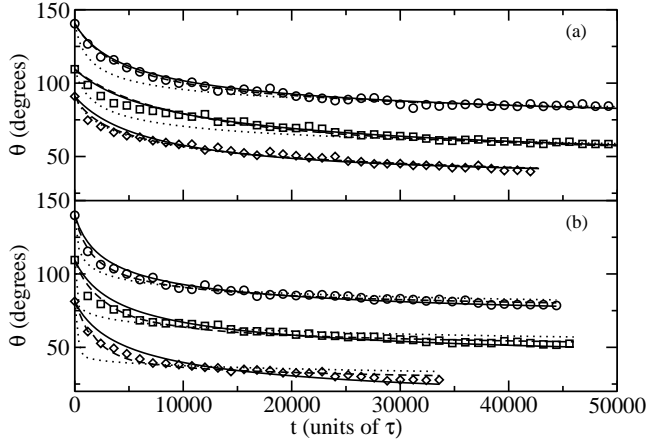}
\caption{\label{cap:cylmodel}(a) Fit of the kinetic (solid line), hydrodynamic
(dotted line), and combined (dashed line) models to the contact angle
data for the cylindrical geometry obtained from MD simulation for
$N=10$ ($\circ$), $N=40$ ($\square$), and $N=100$ ($\diamond$)
droplets for $\varepsilon_{w}=2.0\,\varepsilon$, $\gamma_{L}^{s}=10.0\,\tau^{-1}$.
(b) Same as above with $\varepsilon_{w}=3.0\,\varepsilon$, $\gamma_{L}^{s}=3.0\,\tau^{-1}$.
For clarity, the $N=10$ and $N=40$ data sets have been shifted upward
$60^{o}$ and $30^{o}$, respectively.}

\end{figure}

The fits to the simulation data for various chain lengths for $\varepsilon_{w}=2.0\:\varepsilon$
and $\gamma_{L}^{s}=10.0\,\tau^{-1}$ are shown in Fig. \ref{cap:cylmodel}a.
Figure \ref{cap:cylmodel}b shows results for $\varepsilon_{w}=3.0\:\varepsilon$
and $\gamma_{L}^{s}=3.0\,\tau^{-1}$. The fitting parameters and $\chi^{2}$
values for all of these droplets are listed in Table \ref{cap:paramtable}.
Overall, the combined model produces the best fits to the data, primarily
due to the fact that it has two fitting parameters while the other
two models each have one. The kinetic model also fits the data quite
well in most cases, which suggests that the combined model overspecifies
the droplet spreading behavior for these cases. This is reinforced
by the fact that the parameters extracted from the combined model
do not correspond well with the physical system. In most cases, the
hydrodynamic cutoff $a$ obtained from the combined model is larger
than the droplet radius and the friction coefficient is larger than
that obtained from the kinetic model. Previously \cite{HGW:PRE:03},
the single chain diffusion constant was obtained from simulations
of polymer melts and the friction coefficient was extracted using
the Rouse model via $D=k_{B}T/mN\zeta_{R}$. As for the spherical
droplets, the friction coefficient $\zeta_{0}$ obtained from the
kinetic model was consistently larger than $\zeta_{R}$ for all cases.

\begin{figure}

\includegraphics[%
  clip,
  width=8.5cm,
  keepaspectratio]{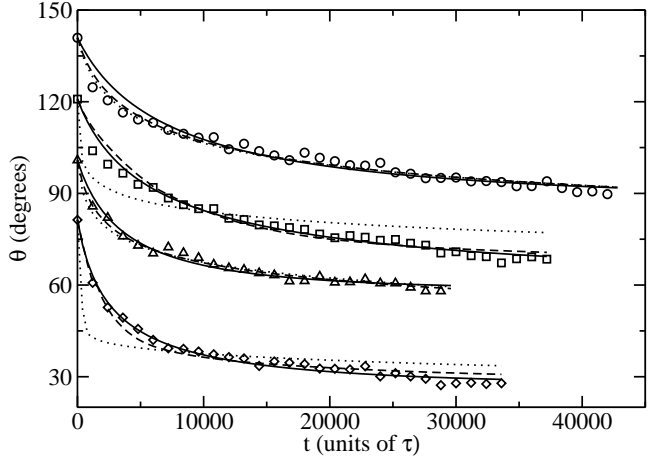}
\caption{\label{cap:model100}Fit of the kinetic (solid line), hydrodynamic
(dotted line), and combined (dashed line) models to the contact angle
data for the cylindrical geometry obtained from MD simulation for
$N=100$ with $\varepsilon_{w}=2.0\,\varepsilon$, $\gamma_{L}^{s}=10.0\,\tau^{-1}$
($\circ$), $\varepsilon_{w}=2.0\,\varepsilon$, $\gamma_{L}^{s}=3.0\,\tau^{-1}$
($\square$), $\varepsilon_{w}=3.0\,\varepsilon$, $\gamma_{L}^{s}=10.0\,\tau^{-1}$
($\triangle$), and $\varepsilon_{w}=3.0\,\varepsilon$, $\gamma_{L}^{s}=3.0\,\tau^{-1}$
($\diamond$). For clarity, the first three data sets have been shifted
upward by $60^{o}$, $40^{o}$, and $20^{o}$, respectively.}

\end{figure}

Although the hydrodynamic model performs better for the cylindrical
geometry than for the spherical geometry and gives values for $a$
that are less than the droplet radius in every case, it still provides
the poorest fit to the data of the three models. As seen from Fig.
\ref{cap:cylmodel}, the best fit to the hydrodynamic model is for
the system with the highest viscosity, $N=100$. To explore this point
in greater detail, we ran two additional systems, $\varepsilon_{w}=2.0\,\varepsilon$
with $\gamma_{L}^{s}=3.0\,\tau^{-1}$ and $\varepsilon_{w}=3.0\,\varepsilon$
with $\gamma_{L}^{s}=10.0\,\tau^{-1}$, for $N=100$. The comparison
of the fits of the three models to all of the $N=100$ systems is
shown in Fig. \ref{cap:model100}. Only for the droplets with the
strongest surface dissipation, $\gamma_{L}^{s}=10.0\,\tau^{-1}$,
does the hydrodynamic model fit the data very well. The hydrodynamic
model fit is very poor for $\gamma_{L}^{s}=3.0\,\tau^{-1}$ regardless
of the equilibrium contact angle. The strong surface dissipation slows
the surface adsorption rate allowing the hydrodynamic behavior to
develop in the bulk region of the droplet. On an atomic substrate,
this is equivalent to increasing the surface corrugation. The combined
model gives fitting parameters that are comparable to both the kinetic
and hydrodynamic models only for the two cases where hydrodynamics
are important. It should be mentioned that the parameters obtained
from the combined model are more sensitive to the input parameters
for the $N=100,$ $\gamma_{L}^{s}=10.0\,\tau^{-1}$ systems. For these
two systems, a 10\% change in the viscosity results  in a 60\% change
in $\zeta_{0}$ and a 15\% change in $a$ for the combined model.
For the other systems, a 10\% change in the viscosity results in a
7\% change in $\zeta_{0}$ and a 1\% change in $a$ on average. The
kinetic model fits the data better for $\varepsilon_{w}=2.0\,\varepsilon$
than $\varepsilon_{w}=3.0\,\varepsilon$, presumably because the driving
force is smaller since the initial droplet is closer to its final
equilibrium contact angle. However, unlike the combined model, the
friction coefficients for these two droplets are consistent with those
for the other droplets so we consider the fit parameters to be accurate.

\subsection{Binary droplets\label{sub:Binary-droplets}}

\begin{figure}

\includegraphics[%
  clip,
  width=8.5cm]{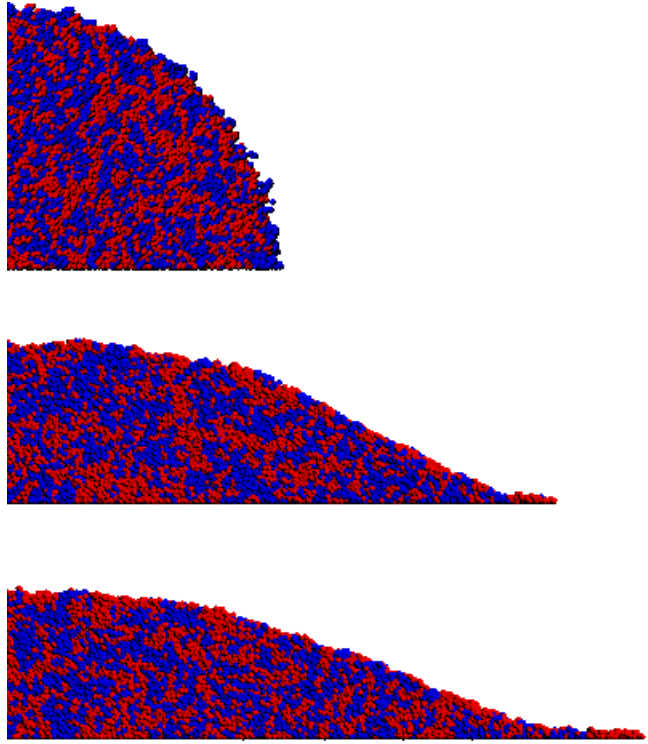}
\caption{\label{cap:snapshot} (Color Online) Snapshots of the spreading binary
droplet containing $352\,000$ monomers of a mixture of $N=10$ and
$N=40$ polymers. The images are taken from time $t=0$ (top), $t=40\,000\,\tau$
(middle), and $t=80\,000\,\tau$ (bottom). Monomers from $N=10$ chains
are shown in red and monomers from $N=40$ chains are shown in blue.
$\varepsilon_{w}=2.0\,\varepsilon$, $\gamma_{L}^{s}=10.0\,\tau^{-1}$.}

\end{figure}

For binary droplets, the behavior is complicated by the interdiffusion
of the two components. As the droplet spreads, the component with
the smaller surface tension gradually diffuses to the droplet surface.
This is evident in the profiles of the binary droplet shown in Fig.
\ref{cap:snapshot}. The droplet, composed of an initial equimolar
mixture of monomers belonging to chains of length $N=10$ and $N=40$,
is on a surface with $\varepsilon_{w}=2.0\:\varepsilon$. Here, the
precursor foot pulls ahead of the bulk region as the droplet spontaneously
wets the surface. The composition of the precursor foot shows a slight
enrichment of the $N=10$ chains, ranging from 59\% to 63\% in the
three frames shown. For $\varepsilon_{w}=3.0\:\varepsilon$, no segregation
in the precursor foot is observed. This can be understood in terms
of the relative distance from the wetting transition for each of the
components. As shown in Fig. \ref{cap:eqangle}, $N=40$ is quite
close to the wetting transition for $\varepsilon_{w}=2.0\,\varepsilon$
and has a significantly slower spreading rate than $N=10$. However,
for $\varepsilon_{w}=3.0\,\varepsilon$, both $N=40$ and $N=10$
are far from the wetting transition and both have fast spreading rates.

\begin{figure}

\includegraphics[%
  clip,
  width=8.5cm]{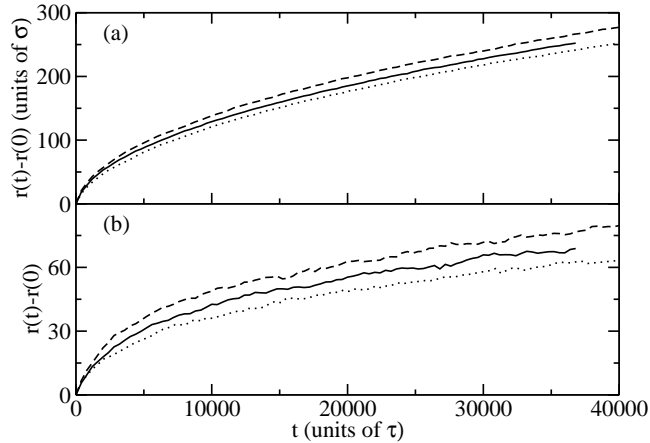}
\caption{\label{cap:blendpure3-40} Spreading rate of (a) the precursor foot
and (b) bulk droplet radius for a $352\,000$ monomer mixture of $N=10$
and $N=40$ polymers compared to homogeneous polymer droplets of the
same size. The curves correspond to $N=10$ (dashed line), $N=40$
(dotted line) and the mixture (solid line). $\varepsilon_{w}=3.0\:\varepsilon$,
$\gamma_{L}^{s}=3.0\,\tau^{-1}$.}

\end{figure}

Figure \ref{cap:blendpure3-40} shows the dynamics of the contact
radius of the precursor foot and the bulk droplet for the $N=10$
and $40$ droplets as well as the droplet consisting of an equal monomer
mole fraction mixture of polymers of chain length $N=10$ and $40$
for $\varepsilon_{w}=3.0\:\varepsilon$ and $\gamma_{L}^{s}=3.0\,\tau^{-1}$.
In general, the spreading rate of the blend falls between that of
the two pure droplets. For either the foot or the bulk, it does not
appear that the dynamics of the blend are being dominated by either
the $N=10$ or $N=40$ polymers. Similar results are found for $\varepsilon_{w}=2.0\:\varepsilon$
and $\gamma_{L}^{s}=10.0\,\tau^{-1}$. This is in agreement with earlier
simulations \cite{VRC:Lan:00,VC:AM:00} of blends of chain length
$N=8$ and $N=16$ where no significant chain length effects are observed. 

\begin{figure}

\includegraphics[%
  clip,
  width=8.5cm]{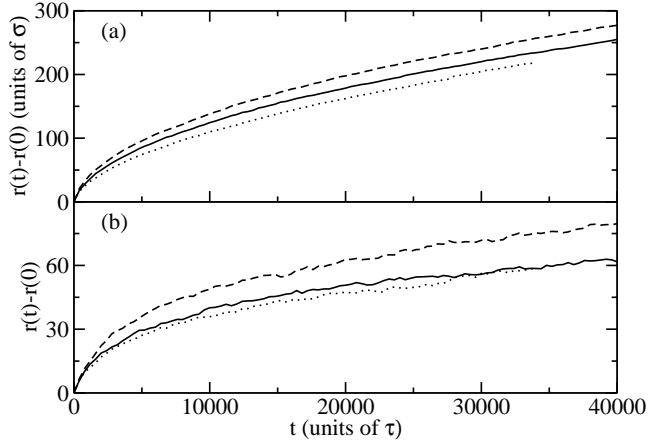}
\caption{\label{cap:blendpure100} Spreading rate of (a) the precursor foot
and (b) bulk droplet radius for a mixture of $N=10$ and $N=100$
polymers compared to homogeneous polymer droplets. The curves correspond
to $N=10$ (dashed line), $N=100$ (dotted line) and the mixture (solid
line). $\varepsilon_{w}=3.0\,\varepsilon$, $\gamma_{L}^{s}=3.0\,\tau^{-1}$.}

\end{figure}

The spreading rate of a droplet consisting of an equal monomer mole
fraction mixture of chains of length $N=10$ and $100$ is shown in
Fig. \ref{cap:blendpure100} for the case where both components are
in the completely wetting regime, $\varepsilon_{w}=3.0\,\varepsilon$.
Here, the contact radius of the bulk region follows the behavior of
the $N=100$ droplet more closely than the $N=10$ droplet. This indicates
that the spreading of the bulk region of the droplet is limited by
the diffusion rate of the larger component. The spreading rate of
the foot is nearly equal to the average spreading rate of the two
pure components and the precursor foot composition is about 50\% short
chain monomers, the same as in the bulk. As found for the $10/40$
blend, there is no measurable segregation for the more energetic surface.
The case where one component wets the surface and the other does not
will be examined in more detail in another paper where we also study
the effect of varying $\varepsilon_{w}$ for two components with the
same chain length. 

The three spreading models are fit to the contact angle data for a
blend of $N=10$ and $N=40$ polymers and for a blend of $N=10$ and
$N=100$ polymers. The fitting parameters and associated errors are
given in Table \ref{cap:binarytable}. In every case, $\zeta_{0}$
from the kinetic model is between that of the corresponding pure component
systems. The hydrodynamic model gives a value for $a$ that agrees
very well with the pure component system of the larger chain length.
The 10/100 mixture is fit much better by the combined model than by
either the kinetic or hydrodynamic models, possibly due to a mixture
of slow and fast dynamics from the two chain lengths.

\section{Velocity distribution\label{sec:Velocity-distribution}}

To analyze the droplet spreading dynamics in greater detail, we consider
the velocity distribution inside the droplet and along the precursor
foot. The velocity at a given position in the droplet is obtained
by generating a histogram of the instantaneous velocities of the monomers.
To eliminate the random fluctuations of the atomic velocities, we
average 500 such histograms over a period of $50\,\tau$. After generating
the averaged histogram, bins containing less than 50 monomers are
manually removed. Otherwise, these nearly empty bins would create
the illusion of more flow at the surface than is actually present. 

\begin{figure}

\includegraphics[%
  clip,
  width=8.5cm]{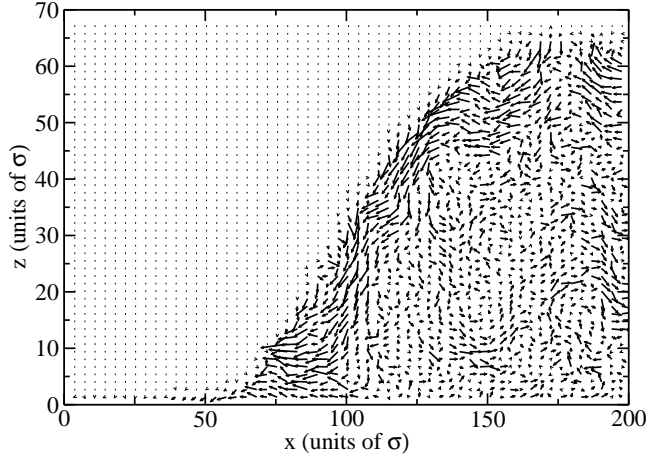}
\caption{\label{cap:mom40} Instantaneous velocity distribution of a homogeneous
droplet after $6000\,\tau$. $N=40$, $\varepsilon_{w}=3.0\,\varepsilon$,
$\gamma_{L}^{s}=3.0\,\tau^{-1}$.}

\end{figure}

The instantaneous velocity of a homogeneous droplet composed of $N=40$
polymers is shown in Fig. \ref{cap:mom40}. The region near the edge
of the droplet has the highest velocity while the monomers at the
center of the droplet and near the substrate but not near the edge
are almost stationary. This is in sharp contrast to previously published
velocity fields of spreading droplets \cite{BCC:CSA:99,RBC:Lan:99}
which show most of the monomers in the droplet moving at the same
speed. In these previous cases, the droplets started as equilibrated
spheres placed just above the substrate, having a contact angle of
$180^{o}$. Thus, these droplets gain a significant amount of momentum
as they move towards the substrate and evolve into the cap geometry
used as the starting point in our simulations. By starting with the
droplet equilibrated at a contact angle of less than $90^{o}$, we
are able to focus on the dynamics induced by the surface tension driving
force and not by the momentum gained by the droplet coming into contact
with the surface.

\begin{figure}

\includegraphics[%
  clip,
  width=8.5cm,
  keepaspectratio]{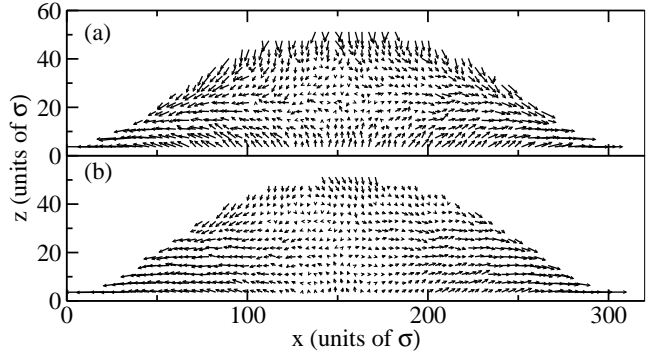}
\caption{\label{cap:mom10-100} (a) Velocity distribution of $N=10$ polymers
in a 10/100 binary droplet after $20\,000\,\tau$. (b) Velocity distribution
of $N=100$ polymers in the same droplet. $\varepsilon_{w}=3.0\,\varepsilon$,
$\gamma_{L}^{s}=3.0\,\tau^{-1}$.}

\end{figure}

The velocity distribution of each component of an equimolar blend
of $N=10$ and $N=100$ polymers is shown in Fig. \ref{cap:mom10-100}.
Instead of averaging the instantaneous velocity over a short period,
the velocity is calculated from the difference in monomer positions
at $20\,000\,\tau$ and $40\,000\,\tau$. This more clearly demonstrates
the slight differences in spreading behavior for the two chain lengths.
Figure \ref{cap:mom10-100} shows that the shorter chains move more
rapidly than the longer chains, but generally exhibit the same behavior.
The shorter chains that are buried in the droplet near the substrate
show a strong tendency to move away from the substrate whereas the
longer chains show no such tendency even though the shorter chains
are farther from the nonwetting transition than the longer chains.
This is possibly due to the ability of the shorter chains to diffuse
a detectable amount during the $20\,000\,\tau$ time period while
the longer chains are considerably slower. Figure \ref{cap:mom10-100}
also indicates that the radial component of the velocity increases
as the distance from the center of the droplet increases. A more detailed
analysis of the source material of the precursor foot shows that the
radial distance traveled by a polymer depends mostly on the initial
radial position of the polymer and not on how near it is to the droplet
surface. Although the highest velocities are found at the droplet
surface, this does not have much influence on the composition of the
precursor foot.

\section{Conclusions\label{sec:Conclusions}}

Using molecular dynamics simulation, we study the spreading dynamics
of binary polymer droplets on a flat substrate. We apply a cylindrical
geometry and demonstrate that the same qualitative spreading behavior
is observed as in a spherical geometry. We derive spreading models
for the cylindrical geometry based on hydrodynamic and kinetic dissipation
mechanisms and show that hydrodynamic effects become relevant at shorter
time scales in the cylindrical geometry. We also show that the $r(t)\sim t^{1/10}$
scaling from the hydrodynamic model and the $r(t)\sim t^{1/7}$ scaling
from the kinetic model for spherical droplets become $r(t)\sim t^{1/7}$
and $r(t)\sim t^{1/5}$, respectively, in the cylindrical geometry.

Fitting the models to the spreading data of homogeneous droplets shows
that the best fit is obtained using a combination of the kinetic and
hydrodynamic dissipation mechanisms, although the parameters extracted
from the fit do not agree well with the physical parameters of the
system. The hydrodynamic model fit the data well only for the slowest
spreading and highest viscosity ($N=100$) case studied, indicating
that the kinetic dissipation mechanism may be dominating any hydrodynamic
effects for the other cases. By increasing the strength of the surface
dissipation, we are able to slow the droplet spreading rate enough
for the $N=100$ for hydrodynamic dissipation to be significant.

Compared to homogeneous droplets, the spreading of binary droplets
is characterized by the difference in surface tension, viscosity,
and interaction strength between the two components and with the substrate.
In the binary droplet at equilibrium, the fraction of shorter chains,
which have a lower surface tension, is higher at the droplet surface.
However, the interdiffusion rate is much slower than the spreading
rate for the droplets presented here. As a result, no enrichment of
the lower surface tension component is observed either at the droplet
surface or in the precursor foot when both components wet the surface.
In the case that the difference in viscosity between the two components
is large, the spreading rate of the bulk region is limited by the
spreading rate of the more viscous component. Otherwise, the spreading
rate is roughly equal to the average rate of the two components. The
single binary system for which the combined model performed noticeably
better than either the kinetic or hydrodynamic models was the mixture
of chain lengths $N=10$ and $N=100$, possibly due to the combination
of fast dynamics from $N=10$ chains and slow dynamics from $N=100$
chains.

By starting with droplets that have a contact angle $\theta=90^{o}$
and not with spherical droplets above the substrate, we are able to
focus on the dynamics induced by the driving forces of droplet spreading
and not by the momentum gained by coming into contact with the substrate.
The instantaneous velocity distribution of the spreading droplet shows
that spreading occurs by the motion of the droplet surface while the
interior of the droplet is almost stationary. For a droplet composed
of an equimolar mixture of short-chain and long-chain polymers, we
find that the shorter chains move more rapidly than the longer ones
near the surface of the droplet. From the droplet surface, they move
downward to the precursor foot and then outward along the substrate. 

Future work will include studying the effects of the interaction strength
between the two components and the effects of patterned surfaces.

\section*{Acknowledgements}

Sandia is a multiprogram laboratory operated by Sandia Corporation,
a Lockheed Martin Company, for the United States Department of Energy's
National Nuclear Security Administration under Contract No. DE-AC04-94AL85000.

\bibliographystyle{apsrev}
\bibliography{cyldrop}

\end{document}